\documentclass{aa}
\usepackage{graphicx}
\begin{document}

\title{The complex galaxy cluster Abell 514:\\
New results obtained with the XMM - Newton satellite
\thanks{Based on observations obtained with \textit{XMM--Newton}, an
ESA science mission with instruments and contributions directly
funded by ESA member states and NASA.}}

\titlerunning{Abell 514: X-ray properties as seen with \textit{XMM--Newton}}

\author{J. Weratschnig\inst{1}
  \and M. Gitti\inst{2}
  \and S. Schindler\inst{1}
  \and K. Dolag\inst{3}}

\offprints{J. Weratschnig, \email{julia.weratschnig@uibk.ac.at}}

\institute{Institut f\"ur Astro- und Teilchenphysik
            Universit\"at Innsbruck, Technikerstrasse 25,
            A-6020 Innsbruck - AUSTRIA\\
            \email{Julia.Weratschnig@uibk.ac.at}
            \and
            INAF - Osservatorio Astronomico di Bologna, via Ranzani 1, 40127
            Bologna - ITALY\\
             \email{myriam.gitti@oabo.inaf.it}
             \and
             Max-Planck-Institut f\"ur Astrophysik, Karl-Schwarzschild-Str.
             1, Postfach 1317, D-85741 Garching
             \email{kdolag@mpa-garching.mpg.de}
             }

\date{}

\abstract{}{We study the X-ray morphology and dynamics of the galaxy
cluster Abell 514. Also, the relation between the X-ray properties
and Faraday Rotation measures of this cluster are investigated in
order to study the connection of magnetic fields and the
intra-cluster medium.}{We use two combined \textit{XMM--Newton}
pointings that are split into three distinct observations.}{The data
allow {us} to evaluate the overall cluster properties like
temperature and metallicity with high accuracy. The cluster has a
temperature of 3.8$\pm$0.2 keV and a metallicity of 0.22 $\pm$0.07
in solar units. Additionally, a temperature map and the metallicity
distribution are computed, which are used to study the dynamical
state of the cluster in detail. Abell 514 represents an interesting
merger cluster with many substructures visible in the X-ray image
and in the temperature and abundance distributions. These results
are used to investigate the connection between the ICM properties
and the magnetic field of the cluster by comparing results from
radio measurements. The new \textit{XMM--Newton} data of Abell 514
confirm the relation between the X-ray brightness and the sigma of
the Rotation Measure ($S_{\rm X}$ - $\sigma_{\rm RM}$ relation)
proposed by Dolag et al. (2001).} {}


\maketitle

\section{Introduction}

It is now well accepted that the intra-cluster medium (ICM)in
clusters of galaxies is magnetized. The magnetic fields can be
traced by diffuse cluster wide synchrotron radio emission
(Giovannini et al. 1991, 1993, Feretti 1999 and Feretti \&
Giovannini 2007) or Inverse Compton hard X-ray radiation caused by
relativistic electrons. Additionally, an indirect measure of the
strength of magnetic fields is the rotation measure (RM), in which
radiation from background radio sources is studied: according to the
strength of the magnetic field inside the cluster, the polarization
angle of the radio emission is rotated. The different observations
lead to the conclusion that magnetic fields in clusters of galaxies
have
strengths of a few $\mu$G (Carilli \& Taylor 2002).\\
Dolag et al. (2001) showed that a relation exists between the X-ray
{\bf surface brightness} and the root mean square scatter
($\sigma_{\rm RM}$) of the Faraday Rotation Measures ($S_{\rm X}$ -
$\sigma_{\rm RM}$ relation) that are used to evaluate the strength
of the magnetic field. This relation is an important tool to study
the connection between the magnetic field and the intra-cluster gas
density and temperature (Dolag et al. 2001). In particular clusters
with polarized extended radio sources are of interest, because it is
possible to evaluate the RM scatter well. More sources in one
cluster give the possibility to get values for the magnetic field
strength in different parts of the cluster, and are therefore very
important observational objects to understand the relation between
the magnetic field and the X-ray properties. In order to compare the
magnetic field and other cluster properties at the position of each
radio source an X-ray image is required. The surface brightness
$S_{\rm X}$ and the RMS
can be determined at the position of each radio source.\\
Since Abell 514 has several radio sources that offer the possibility
to study the $S_{\rm X}$ -  $\sigma_{\rm RM}$ relation, it was chosen for our study. In
this paper, we present results from three \textit{XMM--Newton}
observations
of this cluster. \\
Throughout the paper, a $\Lambda$CDM ($\Omega_{\Lambda}$ = 0.7 and
$\Omega_{\rm m}$ = 0.3) cosmology with a Hubble constant of 70 km
s$^{-1}$ Mpc$^{-1}$  was assumed.

\subsection{Connection of the magnetic field and the ICM density}

The two observables $S_{\rm X}$ and the RMS scatter ($\sigma_{\rm
RM}$) compare the two line of sight integrals:
\begin{equation}
\label{integrals}
    S_{\rm X} \propto \int{n^{2}_{\rm e}\sqrt{T}dx} \leftrightarrow
    \sigma_{\rm RM} \propto \int{n_{\rm e} B_{\|}dx}
\end{equation}
where n$_{\rm e}$ is the electron density and $B_{\|}$ the magnetic
field component parallel to the line of sight. (Dolag et al. 2001;
Clarke et al. 2001)

\begin{figure}[h]
{\includegraphics[width=\columnwidth]{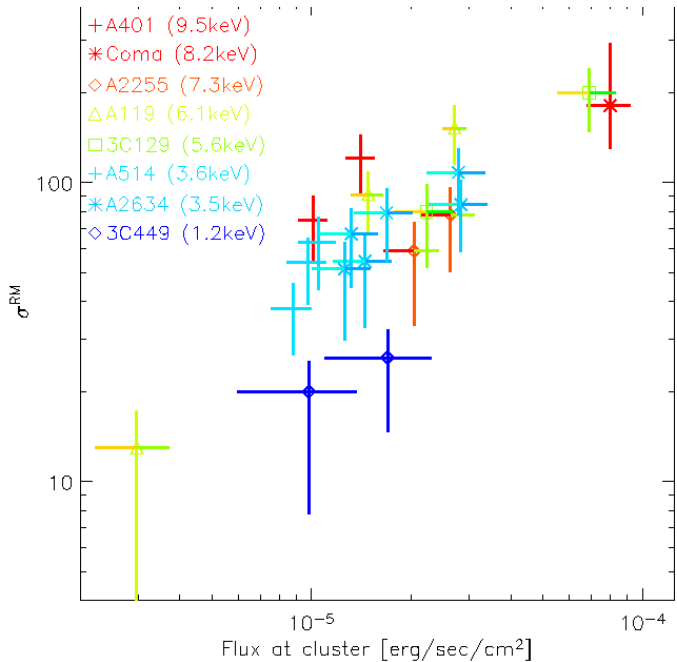}} \caption{The
scatter of the root mean square of the Faraday Rotation measure
(RMS) against the X-ray flux of a sample of clusters, for which both
measurements are available.} \label{rmvsflux}
\end{figure}

When $\sigma_{\rm RM}$ is plotted versus the X-ray flux a clear
relation can be seen. This relation can be fitted by:
\begin{equation}
\label{sigma_rm_relation}
\sigma_{\rm RM} = A\Big(\frac{S_{\rm X}}{10^{-5} \mbox{erg}/\mbox{cm}^2/\mbox{s}}
\Big)^{\alpha}
\end{equation}

A simple interpretation of this relation (e.g. assuming the
temperature within the ICM and the scale-length of the magnetic
field to be fixed) is that the slope $\alpha$ reflects the scaling
of the magnetic field strength ($B$) with the electron density
($n_e$). An exact relation between these two scalings, $B$-$n_{\rm
e}$ and $\sigma_{\rm RM}$-$S_{\rm X}$, is derived in Dolag et al.
(2001) assuming a simplified model for galaxy clusters. Note that
the uncertainties in the 3D position of the individual sources
(which are not known) lead to significant uncertainties in the
derived $\sigma_{\rm RM}$ and therefore imprints a substantial
scatter in the scaling relation. In fact this is the largest
contribution to the the error bars we calculate for $\sigma_{\rm
RM}$ (see Dolag et al. 2001 for details).

Additionally, it seems that there is a suspected dependence on the
cluster temperature: clusters with a high overall temperature also
seem to have high $\sigma_{\rm RM}$ values (see Fig.
\ref{rmvsflux}). To study such matter in detail, clusters that
contain radio sources have to be investigated very accurately in
radio and X-rays.

\section{Abell 514}

\begin{figure}[ht]
{\includegraphics[width=\columnwidth]{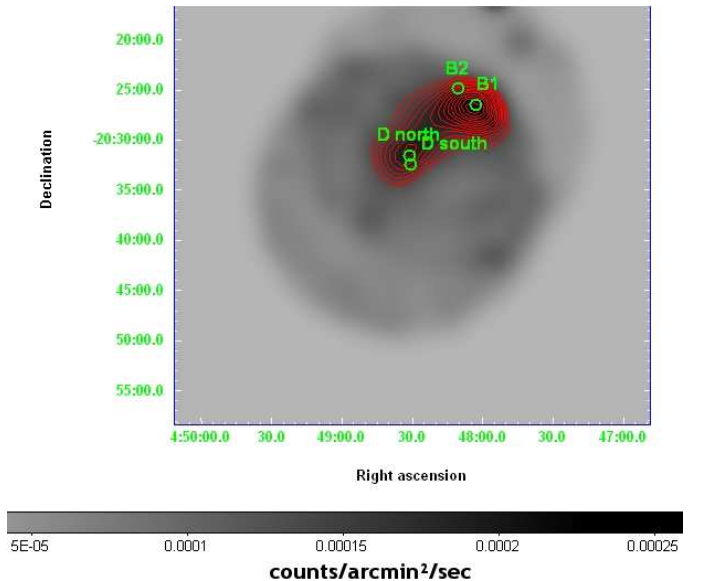}}
\caption{The location of three of the radio sources within the
cluster Abell 514 (D$_{\rm north}$ and D$_{\rm south}$ are
measurements from the same source, but in two slightly offset
positions). The other three radio sources lie outside the field of
view of this X-ray observation.} \label{radio_ps}
\end{figure}

The cluster of galaxies Abell 514 is of Rood-Sastry type F, richness
class 1, and lies between type II and III in the Bautz-Morgan
classification. The cluster was first identified by George Abell
1958 using the National Geographic Society Palomar Observatory Sky
Survey (Abell 1958). In 1966 it was observed by Fomalont \& Rogstad
(1966) during a radio survey at the 21 cm line. Waldthausen et al.
(1979) mapped this cluster using the wavelength $\lambda$ = 11.1 cm.
The optical centre is indicated by Abell et al. (1989) at RA(J2000)
04:47:40 and DEC(J2000) -20:25.7. Earlier X-ray observations were
performed with ROSAT and Einstein
and revealed a highly interesting X-ray morphology (e.g. Govoni et al. 2001).\\
This cluster is very special in several ways. A very prominent
characteristic is the rich morphology that can be seen in ROSAT
images. In contrast to a spherical, relaxed cluster Abell 514 seems
to be in a phase of ongoing merging, making it an example for the
study of dynamical events connected with cluster formation. Another
important point is the fact that six extended radio sources lie
inside the cluster. These radio sources were studied in detail by
Govoni et al. (2001), who derived information on the strength and
structure of the cluster magnetic field by starting from Faraday
Rotation measurements. Three of these sources are within the central
field of view of the {\it XMM--Newton} observations
which we present in this paper. \\
Govoni et al. (2001) found observational evidence for the existence
of a strong magnetic field. The strength of the magnetic field was
estimated to be 4-7 $\mu$G in the centre with a coherence length of
9 kpc. They also give the $\sigma_{RM}$ of the radio sources that
can be seen in the cluster region. Three of them - B2, D North and D
South - (Marked as B2, D north and D south in Fig. \ref{radio_ps})
are inside the field of view of the \textit{XMM} observations and
will be presented in this paper. The radio source B1 was found only
marginally polarized by Govoni et al. (2001)
and is not used as a data point for the $S_{\rm X}$ -  $\sigma_{\rm RM}$ relation.\\

\section{Observations and data reduction}

The data we analyse in this paper result from two different
\textit{XMM--Newton} pointings, split into three distinct
observations. The first observation took place in 2003, February
7th, the second in 2003, March 16th. In 2005, August 15th the
cluster was observed for a third time. All observations were
performed with the European Photon Imaging Camera (EPIC) using the
medium filter in full frame mode. Table
\ref{expTime} displays the exposure times for the individual observations. \\
For the third observation, CCD number six from the MOS 1 camera was
switched off, because of an incident that occurred during revolution
number 961 (the camera was hit by a micrometeroid). Therefore, this
camera is only used for our analysis when the studied area does not
lie inside the affected region.

\begin{table}[h]
\caption{Total and effective exposure times} \centering \label{expTime}
\begin{tabular}{|p{1.7cm}|p{1.8cm}|p{1.8cm}|p{1.8cm}|} \hline
    Camera & Obs. 1 (s) & Obs. 2 (s) & Obs. 3 (s)\\ \hline
    MOS1 tot.&  14963  & 14959  & 15571  \\\hline
    MOS1 eff.& 9388  & 5269  & 5026  \\\hline
    MOS2 tot.& 14963  & 14954  & 15580  \\\hline
    MOS2 eff.& 9355  & 5585  & 5486  \\\hline
    PN tot.&  13388  & 13337  & 14148  \\\hline
    PN eff.& 5007  & 3506  & 3922  \\\hline
\end{tabular}
\end{table}

The data were reduced using SAS version 6.5.

All three observations are heavily polluted by solar flares. The
times with high count rates are therefore rejected.
 The rejection of times with high count rate is done
by creating good time interval tables with defining an upper
threshold for the count rates for each camera and observation. The
times with count rates above the threshold are rejected and new data
sets containing only the flare-free times produced. This threshold
was defined using the count rates in the high energy (10 - 12 keV
for MOS1 and MOS2, 12 - 14 for PN camera) bands. Times where the
count rate was high and also changing with time, were cut out. We
also had a look how the exposure time changes with the threshold:
this curve has at first a steep slope if we take very low thresholds
(cutting away most of the observation time) and gets shallow with
high threshold (cutting away no observation time). A good criterium
to choose the threshold is to take the point where the slope starts
to change. The original and resulting exposure times are listed in
Table \ref{expTime}.\\
To study the diffuse emission of the ICM, point sources are also
removed. This is done by a combination of a source list provided
from the Science Operations Centre (SOC) of XMM data processing and
visual inspection. For each camera and observation, region files
that are to be excluded for the further analysis are created. We
also check if point sources are coincident with the radio sources.
However, this is only the case for B2. For the flux calculation,
the reduced area is taken into account.\\
Also, the images are corrected for the vignetting effect. To achieve
this, we use two different methods. For the image preparation -
especially to get exposure corrected mosaic images - we produce an
exposure map and divide the images by this. Additionally, the method
proposed by Arnaud et al. (2001) is used to correct for vignetting.
Here, every photon is multiplied by a weight factor according to its
position on the detector.\\
Since the PN camera images have many bright columns, they are not
used for the production of a mosaiced and smoothed image. However,
for the spectral analysis, we use the data from those cameras as
well. The areas that show bright pixels or columns are removed by
using a mask.\\
Another important reduction step is the correct
background subtraction. The \textit{XMM} background consists of
three parts (a cosmic X-ray background (CXB), the background
produced by soft proton flares and a non X-ray cosmic background
(NXB) induced by high energy protons). The soft proton flares are
already removed from the data files in the first reduction step,
when the flare free event files are produced. To get rid of the CXB
and the NXB we use the double-subtraction method proposed by Arnaud
et al. (2002) throughout the spectral analysis.

\section{Results}

\subsection{Morphological Analysis}
\label{morph_analysis}

\begin{figure}[h]
{\includegraphics[width=\columnwidth]{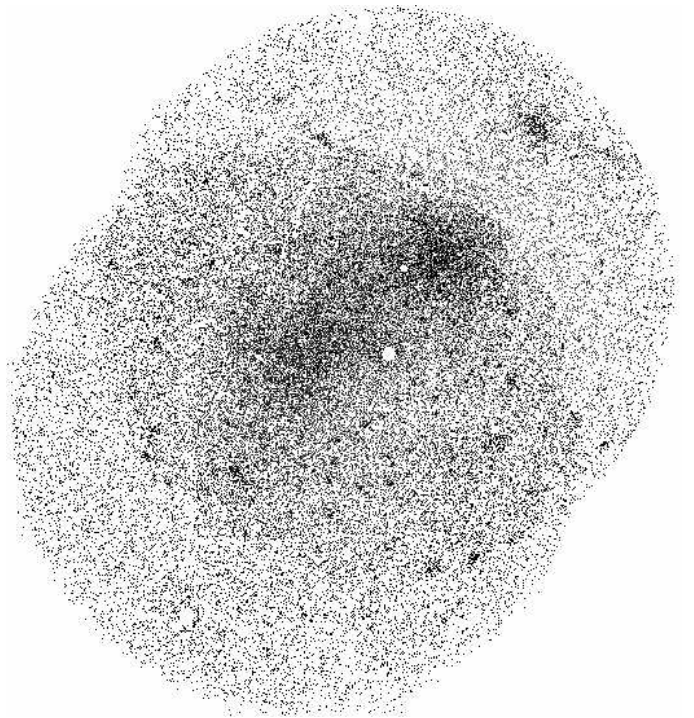}}
\caption{A mosaic image of the MOS cameras of the three different
observations. The image is exposure corrected. The field of view of
the observation is about 37$\times$28 arcmin.} \label{Mosaic}
\end{figure}

\begin{figure}[h]
{\includegraphics[width=\columnwidth]{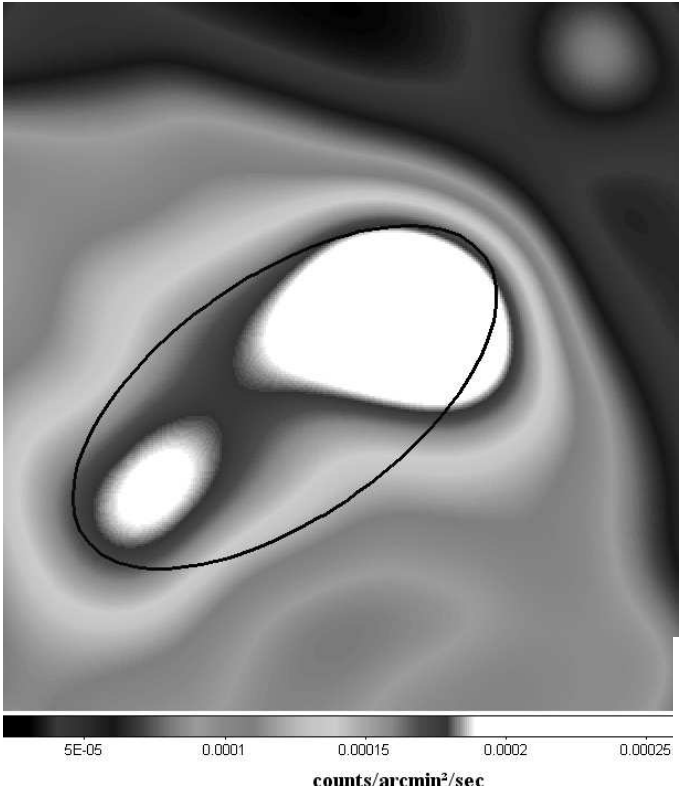}} \caption{A
smoothed image of the whole cluster. This image is corrected for
vignetting and smoothed with an adaptive smoothstyle with a signal
to noise ratio of 40. It shows two subclumps and the overall
elongated shape of the cluster. The size of the box is about 25
arcmin ($\sim$ 2.05 Mpc). The elliptical region indicates the area
where we extracted a spectrum for the whole cluster.} \label{smoIma}
\end{figure}

To study the structure of the cluster in detail, we produce a mosaic
image of the MOS cameras of all three observations using the energy
band between 0.3 and 10 keV (see fig. \ref{Mosaic}). This image is
smoothed using an adaptive smoothstyle and a signal to noise ratio
(SNR) of 40 (see fig. \ref{smoIma}). The adaptive smoothstye is
especially created for poissonian images like X-ray images. Here,
every pixel is assigned with a desired SNR and is then smoothed
towards this SNR by a weighted cyclic convolution. We tried
smoothing the image with different SNR and settled for a SNR of 40,
because with this value the structure of the image is kept and the
borders are not smoothed or enhanced in brightness too much.

The size of the whole field of view of the observation has a length
of 37 and a width of 28 arcmin. This corresponds to a size of 3.0
Mpc x 2.3 Mpc. The ICM emission seems to be elongated along a
filament/main axis over the length of 1.6 Mpc. In the direction
perpendicular to this axis, the cluster emission can be detected out
to 0.8 Mpc. \\

\begin{figure}[h]
{\includegraphics[width=\columnwidth]{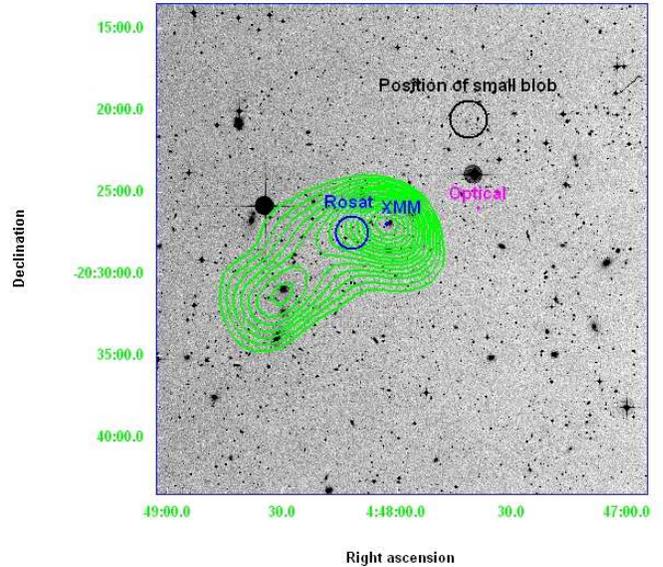}}
\caption{The optical image overlayed with the X-ray contours. The
two main X-ray clumps correspond well with the distribution of the
galaxies, especially the area around the most X-ray bright emission
shows the highest density in galaxies.} \label{opticalXrayCon}
\end{figure}

The X-ray centre lies at RA 04:48:04 (J2000) and DEC -20:26:42
(J2000). The area with the brightest X-ray emission is not a clear
point-like feature. This might be the main reason that this value
differs from the result of earlier observations (Govoni et al.
2001), which give the X-ray centre at RA 04:48:13 (J2000) and DEC
-20:27:18 (J2000). It also depends on the used smoothing method. The
most important point to mention here is however the differently
sized point spread function (PSF) of ROSAT and \textit{XMM--Newton}:
ROSAT's PSF is considerably larger (about 1 arcmin vs. 5-6 arcsec).
This together with the different smoothing methods applied can
explain the offset between the two positions for the X-ray centre.
Especially with a cluster as inhomogeneous as Abell 514 the exact
positioning of a centre is very dependant on smoothing
techniques and detector sensibility.\\
In Fig. \ref{opticalXrayCon} we show the X-ray contours superposed
on an optical image of the cluster (image taken from Aladin
Previewer, Space Telescope Science Institute). The two subclumps
that can be seen in the X-ray image correspond to the galaxy
distribution of the optical image. The X-ray centre is offset with
respect to the optical centre, which is at RA 04:47:40 (J2000) and
DEC -20:25.70 (J2000) (Abell et al. 1989). This offset can be
explained by the fact that Abell 514 is a merger cluster. If we
assume that the Northwest peak has undergone a merger in recent
times (more evidence for this scenario is also discussed in section
\ref{spec_analysis} and \ref{discussion}.) the fact that the galaxy
and gas distributions are offset is not surprising.

The rich substructure that hints at a merger cluster can be seen
clearly in Fig. \ref{smoIma}. To the Northwest of the main cluster a
small blob-like feature is also visible. In the optical image there
are galaxies with cluster redshift seen in the area of this blob.
Therefore we conclude that this is most likely another subpart of
the cluster, which is infalling along the main axis and will merge
with the cluster. It is about 500 kpc away from the closest part of
the rest of the cluster and no connection can be seen towards the
cluster. The brightest peak of the main cluster shows a steeper
decline in surface brightness in the outwards direction than in the
direction towards the second X-ray peak. This feature will be
addressed later (see Sect. 5.1).

\begin{figure}[h]
{\includegraphics[width=\columnwidth]{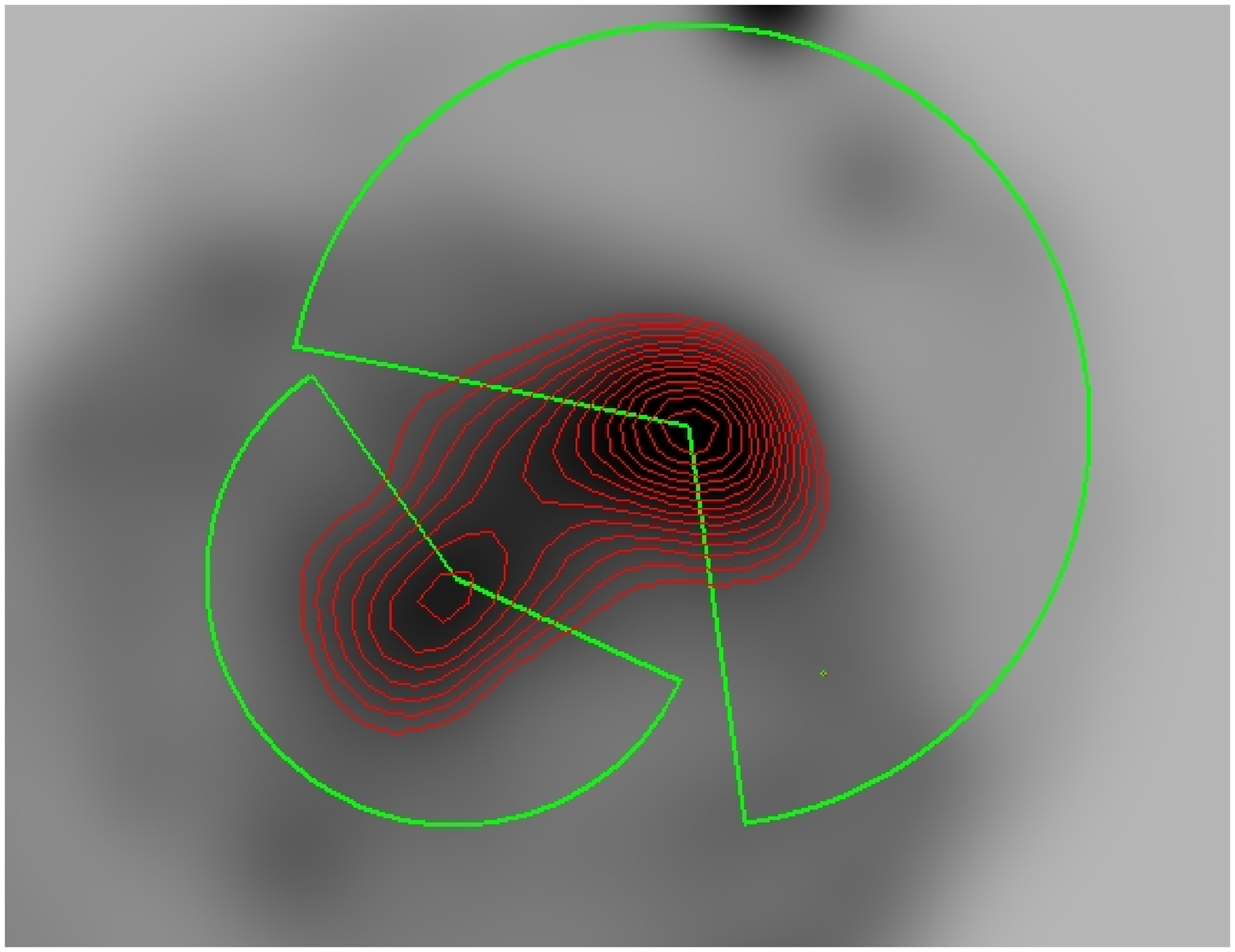}}
\caption{The image shows the regions considered for deriving the
surface brightness profiles around the NW and SE peaks. Each area
was divided in different annuli and the gaps in the detector and
point sources were masked before extracting the surface brightness
profiles.} \label{where_SB}
\end{figure}

\begin{figure}[h]
{\includegraphics[width=\columnwidth]{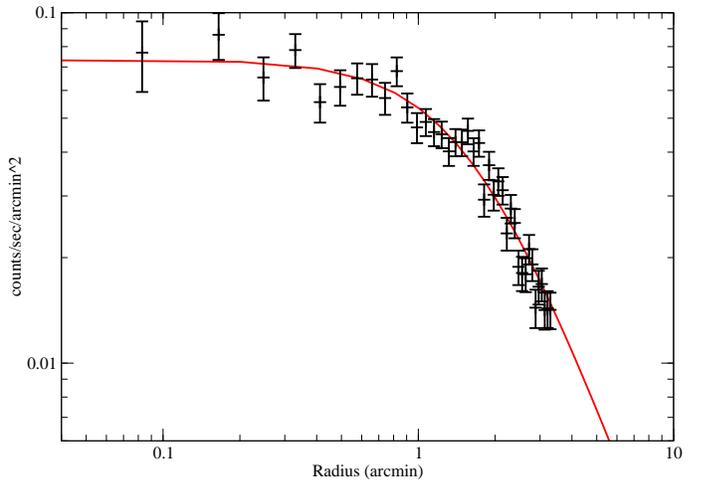}}
\caption{The surface brightness profile around the NW peak (the
X-ray brightest). The profile declines rapidly outside $\sim$ 2
arcmin, a feature most likely caused by a shock due to a merger (see
Sect. 5.1).} \label{NW_peak}
\end{figure}

Around both main peaks visible in the image, the X-ray brightest one
to the Northwest (NW) and the second brightest one to the Southeast
(SE) of the cluster, we extract a surface brightness profile (see
Fig. \ref{where_SB}). In both cases we chose regions that seem to be
mostly unaffected by the merger between those two subparts. To do
this, we selected the areas where no obvious substructures can be
seen in the image (see Fig.\ref{where_SB}). In particular, we
adopted wide-angular regions pointing outwards from the area
connecting the two peaks, where instead substructures can be seen
both in the image and in the temperature map (see Fig.\ref{Tmap}).
To correct for vignetting, a weight factor is applied to the data.
The background is again subtracted using the double background subtraction method.\\
The profile for the NW peak is shown in Fig. \ref{NW_peak}. Apart
from one bump around $\sim$1.5 arcmin from the centre, the profile
around the NW peak does not show any irregularities like bumps or
similar structures. It is noticeable, that the decline between
roughly 1.0 and 2.5 arcmin from the centre is steep compared with a
relaxed cluster. For a relaxed cluster, the surface brightness
profile can be fitted very well with a single $\beta$ profile:
\begin{equation}
\label{betaprofile}
S_{\rm X}(r)=S_{\rm X,0} \Big[1+ \Big( \frac{r}{r_{\rm c}} \Big)^{2}\Big]^{(0.5 - 3\beta)}
\end{equation}

Here, $S_{0}$ is the central surface brightness, $r_{\rm c}$ the
core radius and $\beta$ the slope parameter. In the case of a
relaxed cluster, $\beta$ has a value of roughly 0.6. If we try to
fit the profile of Abell 514 with a single $\beta$ profile, we get a
value of 1.98 for $\beta$. This again shows that it is not a relaxed
cluster part, although no substructure is seen. The steep decline
will be discussed later.

We attempted a similar analysis around the SE peak. We choose five
annuli around the center (see Fig. \ref{where_SB}) in a direction
away from the connection towards the other peak. However, this
analysis was complicated by the low count rates in this region. We
got indications that the surface brightness profile around the SE
peak is shallower than the NW one.

\subsection{Spectral Analysis}
\label{spec_analysis}

As a first step we obtain the temperature and metallicity for the
whole cluster. To get this information, we extract a spectrum in the
elliptical region shown in Fig.\ref{smoIma}. This is done separately
for each camera and observation to maximize the signal to noise
ratio. The background is subtracted using the double subtraction
method proposed by Arnaud et al. (2002). The spectra are then loaded
into Xspec and fitted with a redshifted {\ttfamily MeKaL} model. To
include the Galactic absorption, the Tuebinger
Absorption model ({\ttfamily tbabs}) was used. \\
The energy range for the spectra was between 0.5 and 8.0 keV. This
energy range was chosen because the distinct cameras have the best
agreement in the results in this range. The redistribution matrix
files (RMF) we use are calculated for the MOS cameras using the SAS
task "rmfgen". For the PN camera we adopted the canned matrix
{\ttfamily epn{\_}ff20{\_}sY9{\_}v6.8.rmf}.

\begin{figure}[h]
{\includegraphics[width=\columnwidth]{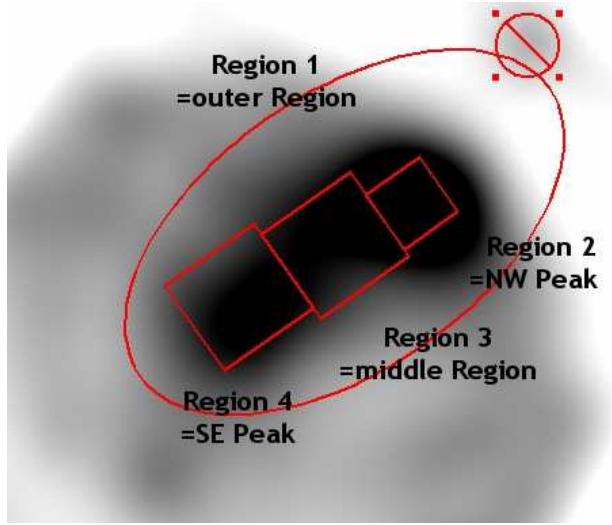}}\caption{The
four regions where temperature and abundance were estimated. The
numbers correspond to the region numbers given in the temperature
and metallicity diagrams.} \label{RegionsTmap}
\end{figure}

The cluster temperature is 3.8 $\pm$ 0.2 keV, which is consistent
with the value of $\sim$ 3.6 keV estimated from the L-T relation
(Govoni et al. 2001). The overall cluster metallicity is 0.22 $\pm$
0.07 in solar units. \footnote{The {\ttfamily MeKaL} fit
gives a reduced $\chi^2 \sim 1.8$.} \\
To study the temperature and metallicity distribution in detail, we
divide the cluster into four regions and extract a spectrum in each
one. This is done for all three observations for all cameras. Again,
the resulting spectra are fitted in Xspec with a {\ttfamily MeKaL}
model. Fig. \ref{RegionsTmap} shows the regions where the spectra
were extracted. The regions are chosen to contain a comparable
photon signal and also give comparable statistics. The region
numbers are defined in the following way: region 1 = outer region,
region 2 = box around NW peak, region 3 = area between the two
peaks, region 4 = box around SE peak. The final values for
temperature and abundance do not change if those areas are moved
around, as long as they cover the area around the NW peak, the
region between the two peaks, the SE peak and the outskirts of the
cluster. With the regions we give here, we are able to collect most
photons per area and get better statistics then e.g. choosing circles as regions.\\

\begin{figure}[h]
{\includegraphics[width=\columnwidth]{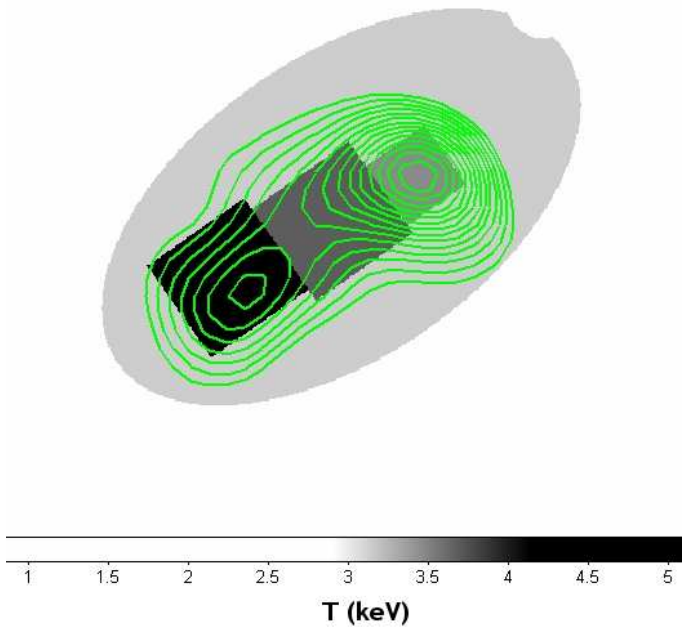}}
\caption{The temperature map of the cluster. Overlayed are the
contours of the X-ray surface map. The circle represents the region
where a subpart of the cluster is visible in the raw and smoothed
images (see Figs. 3 and 5). This area was excluded from the spectral
analysis.} \label{Tmap}
\end{figure}

By comparing the temperature and metallicity distribution we are
able to study the dynamical state of the cluster. In Fig. \ref{Tmap}
the temperature map which is calculated using spectra in different
regions of the cluster is shown. Three regions with different
temperature along the axis of the cluster can be seen, as well as a
cooler outer region.

\begin{figure}[h]
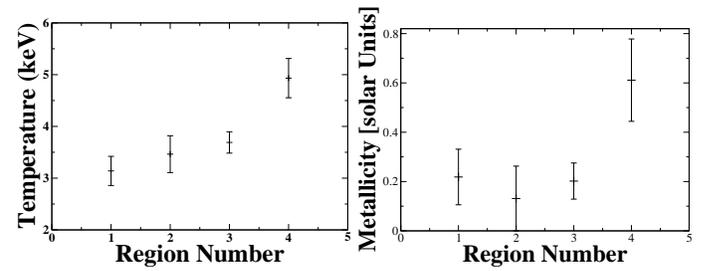

{\includegraphics[width=4.45cm]{TemDistr.eps}}
\phantom{}\includegraphics[width=4.45cm]{AbunDistr.eps}\caption{The
temperature and the metallicity distribution in the cluster. The
region numbers are as defined in Fig. \ref{RegionsTmap}.}
\label{TdisAbunDis}
\end{figure}

The hottest region is the box number four which is located around
the SE peak. It is also the one with the highest metallicity, as can
be seen in the second diagram in Fig. \ref{TdisAbunDis}. The right
panel in Fig. \ref{TdisAbunDis} shows the metallicity distribution
in the cluster. We see that the SE peak has a higher
metallicity than the rest of the cluster.\\
Inside the error bars the temperatures derived for the NW region and
the middle region can be seen as having the same temperature as the
outside region. There is a trend in the cluster to have higher
temperatures in the SE. The region around the SE peak is clearly the
hottest of the whole cluster.

The difference in metallicity between regions two and three compared
to region four, can be seen as a sign that those parts of the
clusters have not yet had the possibility to merge and are still
infalling towards a common centre. As has been shown by Kapferer et
al. (2006), a cluster has steeper gradients in metallicity before
the merger process. When the
subclusters have finally merged, their metallicity is smoothly distributed.\\

\begin{figure}[h]
{\includegraphics[width=\columnwidth]{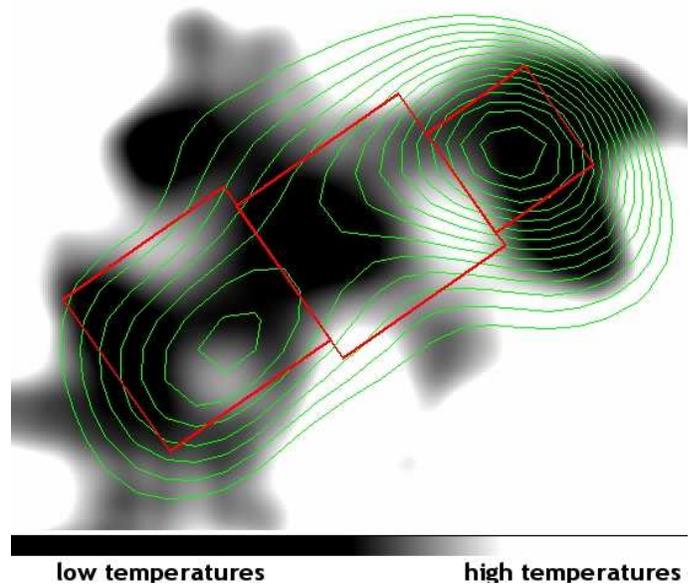}} \caption{The
temperature map created using the hardness ratio of images in four
different energy bands. See text for details.} \label{HRtmap}
\end{figure}

Another way to study the temperature distribution is via
hardness-ratio maps. Such a map is also produced for this cluster
from four different energy bands (0.3-1, 1-2, 2-4.5, 4.5-8 keV).
Only the MOS1 camera of the first observation could be used for this
due to technical reasons. Therefore, the count rates are very low
compared to the other method and only relative differences in
temperature but no absolute values can be shown. The temperature map
is presented to show that the temperature distribution is very
inhomogeneous. This hints at a merger cluster which is not yet
relaxed but in the first stages of
merging (Fig. \ref{HRtmap}). \\

The region of the brightest X-ray peak is cool, which is in good
agreement with the spectral result that also gives a low temperature
for this part of the cluster. The second brightest X-ray peak has a
higher temperature, again corresponding to the spectral result that
gives a higher temperature for the area around this peak. The region
between the two peaks seems to be a mix of high and low
temperatures, corresponding to the mean temperature of the spectral
result. The seperate areas with different temperatures between the
two X-ray clumps cannot be seen using the spectral method, since we
do not have enough photons to produce a spectrum that can be fitted
reliable. We therefore see the mixing of the different temperatures.
The regions in the outer parts of the cluster have too low count
rates to give reliable results.

\subsection{Mass determination}

When we assume hydrostatic equilibrium and spherical symmetry, it is
possible to calculate the mass of a galaxy cluster using the
temperature and the density profiles.
Although Abell 514 is a very active merger cluster and neither in a
hydrostatic equilibrium nor has a spherical shape, we try to use
these assumptions to calculate the mass of two subparts of the
cluster. These two parts are the regions around the two X-ray
brightest peaks. They show a separated emission and can be
approximated as spherical symmetric in a first, rough step.\\
The total mass is given by the equation:

\begin{equation}
\label{hydroEquation}
M_{\rm tot}(r) = -\frac{kT}{G\mu m_{\rm p}}r\Big[\frac{d \ln n_{\rm e}}{d \ln r} +
\frac{d \ln T}{d \ln r}\Big]
\end{equation}

where $k$ is the Boltzmann constant, $T$ the gas temperature, $G$ the
gravitational constant, $\mu$ the mean molecular weight of the gas
($\mu$ $\approx$0.6), $m_{\rm p}$
the proton mass and $n_{\rm e}$ the electron density.\\
If the ICM follows a $\beta$-model, the electron density can be
written as:

\begin{equation}
\label{electronDensity}
n_{\rm e}(r) =
n_{\rm e0}\Big[1+\Big(\frac{r}{r_{\rm c}}\Big)^2\Big]^{-\frac{3}{2}\beta}
\end{equation}

The values for $\beta$ and $r_{\rm c}$ are the values obtained by fitting
a $\beta$ profile to the surface brightness of the cluster.\\
Inserting equation \ref{electronDensity} into equation
\ref{hydroEquation}, yields:

\begin{equation}
\label{masseNonIsothermal}
M_{\rm tot}(r) =-\frac{kr^2}{G\mu m_{\rm p}}\Big[\frac{dT}{dr} - 3\beta
T\frac{r}{r^2+r_{\rm c}^2}\Big]
\end{equation}

Assuming that the cluster is isothermal inside a certain radius,
$\frac{dT}{dr}$ is zero. The final equation to calculate the mass
inside a certain radius is therefore:

\begin{equation}\label{masseIsothermal}
M_{tot}(r) = \frac{3k\beta}{G\mu m_p} T\frac{r^3}{r^2+r_c^2}
\end{equation}

With the values we obtain by trying to fit a single $\beta$ model to
the surface brightness profiles of the two brightest peaks, we are
able to give at least a very rough first estimate of the masses.
Since we can extract the profile of the second brightest peak only
out to 5.7 arcmin ($\sim$ 490 kpc), we use this radius to calculate
the mass for both regions. Using equation \ref{masseIsothermal} and
the results from the spectral analysis for the temperature in the
different parts of the cluster (region 2 and 4, see below) the mass
of the X-ray brightest part inside a radius of $\sim$490 kpc is
about 3.0 10$^{14}$ M$_{\odot}$, while the second clump has a mass
of about 6.5 10$^{13}$ M$_{\odot}$. This can only be seen as a crude
first guess of the masses. The X-ray brightest part also seems to be
the most massive one. This result can be expected from the L$_X$ -
Mass relation.

\section{Discussion}\label{discussion}

\subsection{Candidate for a cold front or a shock?}

A prominent morphological structure of Abell 514 is a steep decline
in X-ray surface brightness towards the Northwest region. This can
be seen as a sharp edge in the image (see Fig. \ref{smoIma}), as
well as a quick drop of the surface brightness profile outside
$\sim$ 2 arcmin (see Fig. \ref{NW_peak}).

\begin{figure}[h]
{\includegraphics[width=\columnwidth]{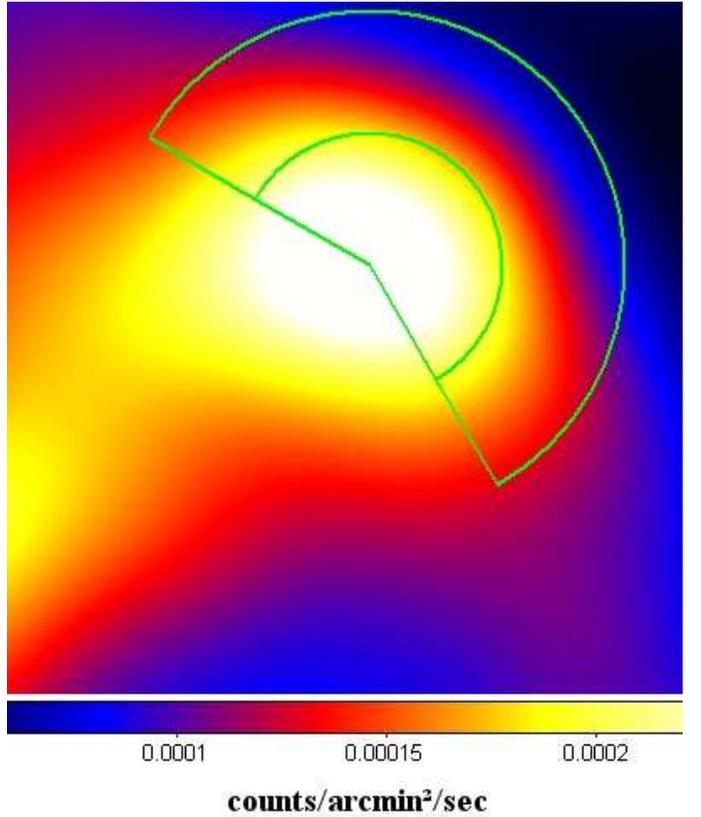}}
\caption{Regions considered for the deprojection analysis in order
to investigate the nature of the drop in surface brightness.}
\label{where_depr}
\end{figure}

Possible explanations for such a feature can be either a cold
front or a shock caused by the merger process.
Similar features were found by Markevitch et al. (2000) and
Vikhlinin et al. (2002) in the clusters Abell 2142 and Abell 3667.
Another example for a similar structure was also found in Abell 2256
by Sun et al. (2002). During a cluster merger, a cool core of a
subpart of a cluster can survive the merging process. This is
characterised by the fact that the temperature inside a brightness edge
is lower than in the surrounding region.
The other explanation for a feature like the one seen in Abell 514
would be a shock where the material is compressed.\\
To test if the edge in Abell 514 is caused by a cold front or a
shock we study two regions, one inside and one outside the edge
visible in the smoothed X-ray image (Fig. \ref{smoIma}), with
respect of their density and temperature. The regions used for this
analysis are shown in Fig. \ref{where_depr}.

Region 1 is the region inside the "edge", while region 2 is the area
in the outer part. We apply the deprojection method by using the
Xspec model {\ttfamily projct} to calculate densities inside and
outside of this border. Also the temperatures in both regions were
calculated and compared with each other. The results are shown in
Table \ref{ColdFront_Shock}.

\begin{table}[h]
\caption{The density and temperature inside (Region 1) and outside
(Region 2) the brightness edge.} \centering \label{ColdFront_Shock}
\begin{tabular}{|c|c|c|} \hline
     & Region 1 & Region 2 \\ \hline
    Density [10$^{-3}$cm$^{-3}$]& 0.91 $\pm$ 0.11 & 0.51 $\pm$ 0.06 \\\hline
    Temperature [keV]& 4.5 $\pm$ 0.8 & 3.6 $\pm$ 0.5 \\\hline
  \end{tabular}
\end{table}

The temperature inside the border is slightly higher than outside,
but no jump in temperature can be deduced from our data, especially
not a jump from a cool core to a warmer surrounding. Inside the
errorbars, both temperatures can be seen as the same. Therefore the
discontinuity in surface brightness cannot be caused by a cold
front. The density however shows a clear discontinuity. It is
therefore possible that the brightness jump is due to a shock. Such
a shock can be the result from an earlier merger, with the
different structures not distinguishable by eye any more. \\
The visible interaction between the SE peak and the NW one is most
likely not responsible for this feature. We see that the
metallicities between the two peaks are very different. It is
therefore plausible that they have not merged yet and cannot cause
the feature in the surface brightness seen in the NW peak.
\\
Another possibility could be an interaction of the main X-ray peak
with the small blob from the south east part. But since this
structure is still 500 kpc away from the main cluster and no
connection between the two parts can be seen we do not expect to see
any interaction effects yet between those parts.\\

\subsection{The $S_{\rm X}$ -  $\sigma_{\rm RM}$ relation and the magnetic field}

According to theory (Tribble 1993, Dolag et al. 1999), the magnetic
field is amplified in a hot merger cluster. The $S_{\rm X}$ -
$\sigma_{\rm RM}$ relation is clearly dependent on the temperature
of the cluster (see Sect. 1.1). For Abell 514, this general trend
can be studied. Although Abell 514 is a merger cluster, its magnetic
field is still quite low. This can be seen in good agreement with
the low overall temperature of the cluster. Still, compared to other
cool clusters, Abell 514 shows a slightly higher $\sigma_{RM}$ which
is most likely due to the ongoing merger that already enhanced the
magnetic
field.\\
One main aim of the \textit{XMM--Newton} observations was to get new
values for the X-ray flux in the regions where the radio sources
are. It has to be mentioned that the true location of these radio
sources inside the cluster is not known. This fact is taken into
account in the errors given for the $\sigma_{\rm RM}$ value. The
error bars cover the range of values between a source located in the
cluster center and one behind the cluster.

\begin{figure}[h]
{\includegraphics[width=\columnwidth]{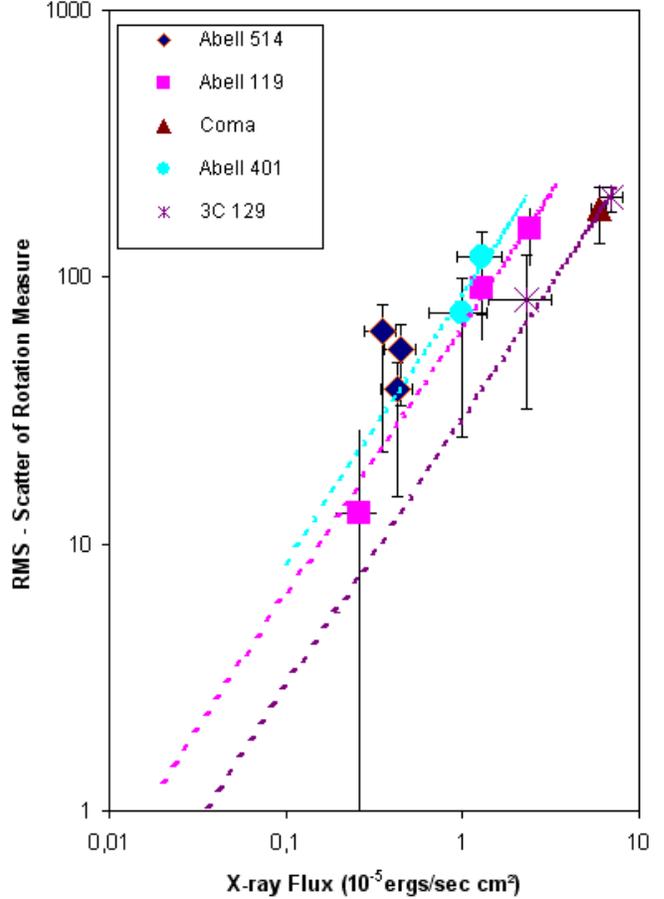}}
\caption{The X-ray flux -  $\sigma_{\rm RM}$ relation with the new
data points for Abell 514. The new values are in good agreement with
the general slope of the relation} \label{RMS_neu}.
\end{figure}

The new values are then compared to the results from measurements of
the magnetic field (via the $S_{\rm X}$ - $\sigma_{\rm RM}$). They
fit well with other measurements from Coma, A119 etc. (see Fig.
\ref{RMS_neu}). Fig. \ref{RMS_neu} shows the results from the new
measurements, with the data point for the other clusters (Coma,
A119, etc.) being converted to the same energy band. In general, the
data points obtained for A514 are in good agreement with the
relation found from the rest of the clusters. The lines in fig.
\ref{RMS_neu} represent the correlations for the distinct clusters.
We see that the data points of Abell 514 lie above the correlations
of all the other clusters. This can be seen as an indication for an
amplification of the magnetic field due to the ongoing merger in
Abell 514. Overall, the points from Abell 514 make the whole
correlation (if we use the observational data) less steep. Without
the data points from Abell 514, the slope parameter is 1.19, while
it is 0.98 with them. Inside an error of 10\% both values agree. To
avoid any instrumental bias in this study in the future, we plan to
obtain {\it XMM--Newton}
data for the other clusters in this sample as well.\\

Additionally, with the creation of a temperature map, it is possible
to compare the strength of the RMS scatter $\sigma_{\rm RM}$ from the
rotation measures with the temperature of the ICM in the area of the
radio source. Table \ref{RMT2} shows the results.

\begin{table}
  \caption{Comparison of $\sigma_{\rm RM}$ with the temperature.
  The regions are as indicated in Fig. \ref{RegionsTmap}}
  \label{RMT2} \centering
  \begin{tabular} {|p{3cm}|c|p{2.3cm}|} \hline
    Region Nr. & T (keV) & $\sigma_{RM} (rad/m^{2})$ \\\hline
    Region 2 (includes Radio source B2) & 3.2 $\pm$ 0.2 & 63 $^{+16}_{-41}$\\\hline
    Region 4 (includes Radio sources D$_{north}$ and D$_{south}$) & 4.9 $\pm$ 0.4 &
    54 $^{+12}_{-21}$ (north) 38 $^{+10}_{-23}$(south) \\\hline
  \end{tabular}
\end{table}

Here, $\sigma_{\rm RM}$ is lower in the hottest region and higher in
the cool, X-ray brightest part, that seems to be the most relaxed
part of the cluster. However, this is not in contradiction with the
above relation. Inside the cluster, more complicated effects take
place additionally to the overall properties, that cannot yet be
resolved with the current observations. Also, the RMS measurements
are taken from a smaller area than the spectra we use to deduce the
temperature. Small scale fluctuations inside these regions are
therefore possible and not taken into account in table \ref{RMT2}.

\section{Summary}

We performed a detailed study of the X-ray emission of the merger
cluster Abell 514. Three pointings by the \textit{XMM--Newton}
telescope were analysed to study the properties of this cluster,
especially the dynamical state and the relation between the X-ray
flux and the RMS of the
rotation measure produced by the magnetic field inside the cluster.\\
The image of Abell 514 shows the rich substructure of the cluster, a
clear sign for an ongoing merger. Two main X-ray bright peaks can be
seen with a connection between them. The brightest peak also shows
signs for a shock, most likely caused by a recent merger.\\
We found the overall cluster temperature to be 3.8 $\pm$ 0.2 keV.
This value is in good agreement with the one from the L-T relation
(3.6 keV). The cluster metallicity is 0.22 $\pm$ 0.07 solar units.\\
Additionally to the calculation of overall values for the
temperature and the metallicity we are able to produce rough
temperature and metallicity maps. To achieve this, we divide the
cluster in four different regions and extracted spectra therein. With
the help of these maps, we can study the dynamical state of the
cluster in more detail.

It appears that the two main visible subclumps have not had time to
merge yet. Their temperatures and metallicities have significantly
different values. The brightest part in the Northeast shows a steep
decline that could be caused by a shock due to an earlier merger. We
divide this area into two regions to calculate the density and
temperature inside and outside the visible edge. The obtained values
indicate that the brightness edge is indeed caused
by a shock.\\
The X-ray flux is determined in the regions where extended radio
sources are. These radio sources enable the measurement of the
scatter of the Faraday Rotation measures which is due to the
strength of the magnetic field. They are related with the X-ray
flux. With the \textit{XMM--Newton} observations we are able to add new
points to this $S_{\rm X}$ - $\sigma_{\rm RM}$ relation. The new data points
fit well in the model predicted by Dolag et al. (2001).\\
The low overall temperature also confirms the relation between the
ICM temperature and the magnetic field strength (lower temperature
clusters have generally smaller magnetic fields). This can also be
seen as a sign that the cluster is still in an early stage of the
merger and has not been heated up yet, nor has the magnetic field
been enhanced by the merger.

\section*{Acknowledgments}
We wish to thank the referee for helpful comments.
We thank E. Pointecouteau for the help with the spectral
temperature map in Fig. \ref{Tmap} and S. Ettori for providing the
software required to produce the hardness ratio map in Fig.
\ref{HRtmap}. We also thank C. Sarazin for fruitful discussions
and help with the topic. M. Gitti acknowledges support by grant
ASI-INAF I/088/06/0. J. Weratschnig thanks the European Science
Foundation (ESF). S. Schindler acknowledges the Austrian Science
Foundation FWF grants P19300-N16 and P18523-N16.


\begin{thebibliography}{} 
\bibliographystyle{aa}


\bibitem[Abell(1958)]{1958ApJS....3..211A} Abell, G.~O.\ 1958, \apjs, 3,
211

\bibitem[Abell et al.(1989)]{1989ApJS...70....1A} Abell, G.~O., Corwin,
H.~G., Jr., \& Olowin, R.~P.\ 1989, \apjs, 70, 1

\bibitem[Arnaud et al.(2001)]{2001A&A...365L..80A} Arnaud, M., Neumann,
D.~M., Aghanim, N., Gastaud, R., Majerowicz, S., \& Hughes, J.~P.\
2001, \aap, 365, L80

\bibitem[Carilli \& Taylor(2002)]{2002ARA&A..40..319C} Carilli, C.~L., \&
Taylor, G.~B.\ 2002, \araa, 40, 319

\bibitem[Clarke et al.(2001)]{2001ApJ...547L.111C} Clarke, T.~E., Kronberg,
P.~P., \& B\"{o}hringer, H.\ 2001, \apjl, 547, L111

\bibitem[Dolag et al.(1999)]{1999A&A...348..351D} Dolag, K., Bartelmann,
M., \& Lesch, H.\ 1999, \aap, 348, 351

\bibitem[Dolag et al.(2001)]{2001A&A...378..777D} Dolag, K., Schindler, S.,
Govoni, F., \& Feretti, L.\ 2001, \aap, 378, 777

\bibitem[Feretti et al.(1999)]{1999A&A...344..472F} Feretti, L., Dallacasa,
D., Govoni, F., Giovannini, G., Taylor, G.~B., \& Klein, U.\ 1999,
\aap, 344, 472

\bibitem[Fomalont \& Rogstad(1966)]{1966ApJ...146..528F} Fomalont, E.~B.,
\& Rogstad, D.~H.\ 1966, \apj, 146, 528

\bibitem[Giovannini et al.(1991)]{1991A&A...252..528G} Giovannini, G.,
Feretti, L., \& Stanghellini, C.\ 1991, \aap, 252, 528

\bibitem[Giovannini et al.(1993)]{1993ApJ...406..399G} Giovannini, G.,
Feretti, L., Venturi, T., Kim, K.-T., \& Kronberg, P.~P.\ 1993,
\apj, 406, 399

\bibitem[Govoni et al.(2001)]{2001A&A...379..807G} Govoni, F., Taylor,
G.~B., Dallacasa, D., Feretti, L., \& Giovannini, G.\ 2001, \aap,
379, 807

\bibitem[Feretti \& Giovannini(2007)]{2007astro.ph..3494F} Feretti, L., \&
Giovannini, G.\ 2007, ArXiv Astrophysics e-prints,
arXiv:astro-ph/0703494

\bibitem[Kapferer et al.(2006)]{2006A&A...447..827K} Kapferer, W., Ferrari, C., Domainko, W., et al.\
2006, \aap, 447, 827

\bibitem[Markevitch et al.(2000)]{2000ApJ...541..542M} Markevitch, M., Gonzalez, A. H.,
 David, L., Vikhlinin, A., et
al.\ 2000, \apj, 541, 542

\bibitem[Sun et al.(2002)]{2002ApJ...565..867S} Sun, M., Murray, S.~S.,
Markevitch, M., \& Vikhlinin, A.\ 2002, \apj, 565, 867

\bibitem[Tribble(1993)]{1993MNRAS.263...31T} Tribble, P.~C.\ 1993, \mnras,
263, 31

\bibitem[Vikhlinin \& Markevitch(2002)]{2002AstL...28..495V} Vikhlinin,s
A.~A., \& Markevitch, M.~L.\ 2002, Astronomy Letters, 28, 495

\bibitem[Waldthausen et al.(1979)]{1979A&AS...36..237W} Waldthausen, H.,
Haslam, C.~G.~T., Wielebinski, R., \& Kronberg, P.~P.\ 1979, \aaps,
36, 237


\end{thebibliography}
\end{document}